\documentclass[preprint,12pt]{elsarticle}

\usepackage{amssymb}

\journal{Acta Astronautica}

\begin{document}

\begin{frontmatter}



\title{EMMI -- Electric Solar Wind Sail Facilitated Manned Mars Initiative}


\author{Pekka Janhunen, 
 Sini Merikallio and Mark Paton}

\address{Corresponding author: P. Janhunen, Finnish Meteorological Institute, 
Erik Palm\'enin aukio 1, FI--00560 Helsinki, Finland.
(pekka.janhunen@fmi.fi)}

\begin{abstract}
The novel propellantless electric solar wind sail concept promises efficient low thrust transportation in the Solar System outside Earth's magnetosphere. Combined with asteroid mining to provide water and synthetic cryogenic rocket fuel in orbits of Earth and Mars, possibilities for affordable continuous manned presence on Mars open up. 
Orbital fuel and water enable reusable bidirectional Earth-Mars vehicles for continuous manned presence on Mars and allow smaller fuel fraction of spacecraft than what is achievable by traditional means. Water can also be used as radiation shielding of the manned compartment, thus reducing the launch mass further. In addition, the presence of fuel in the orbit of Mars provides the option for an all-propulsive landing, thus potentially eliminating issues of heavy heat shields and augmenting the capability of pinpoint landing. With this E-sail enabled scheme, the recurrent cost of continuous bidirectional traffic between Earth and Mars might ultimately approach the recurrent cost of running the International Space Station, ISS.
\end{abstract}


\end{frontmatter}


\section{Introduction}
\label{sec-intro}
Manned missions to Mars have been in the planning stage since the onset of space age~\citep{portree2001, dorney2012}. Current proposals include national space agency plans as well as private sector efforts. 
The European Agency (ESA) as well as the National Aeronautics and Space Administration (NASA) have both expressed interest in manned Mars flights. Private sector ventures, such as MarsOne or Mars Direct have more ambitious schedules, but are struggling with resource limitations and technical feasibility issues~\cite{do2014}. All of the above base the transportation on traditional propulsion including heavy launchers. Moreover, it has been estimated that MarsOne would require 15 Falcon Heavy launches to initiate the project even before manned flight phase~\cite{do2014}. 

The electric solar wind sail (E-sail) \cite{janhunen2010big} is a novel propellantless propulsion concept utilizing the solar wind. It is estimated to be very efficient in terms of impulse versus propulsion system mass \cite{janhunen2013}. As its propulsion system is lightweight and does not consume any propellant, the E-sail can transport cargo payloads in the solar system with reasonable costs and flight times.

\begin{figure}
 \noindent\includegraphics[width=15cm]{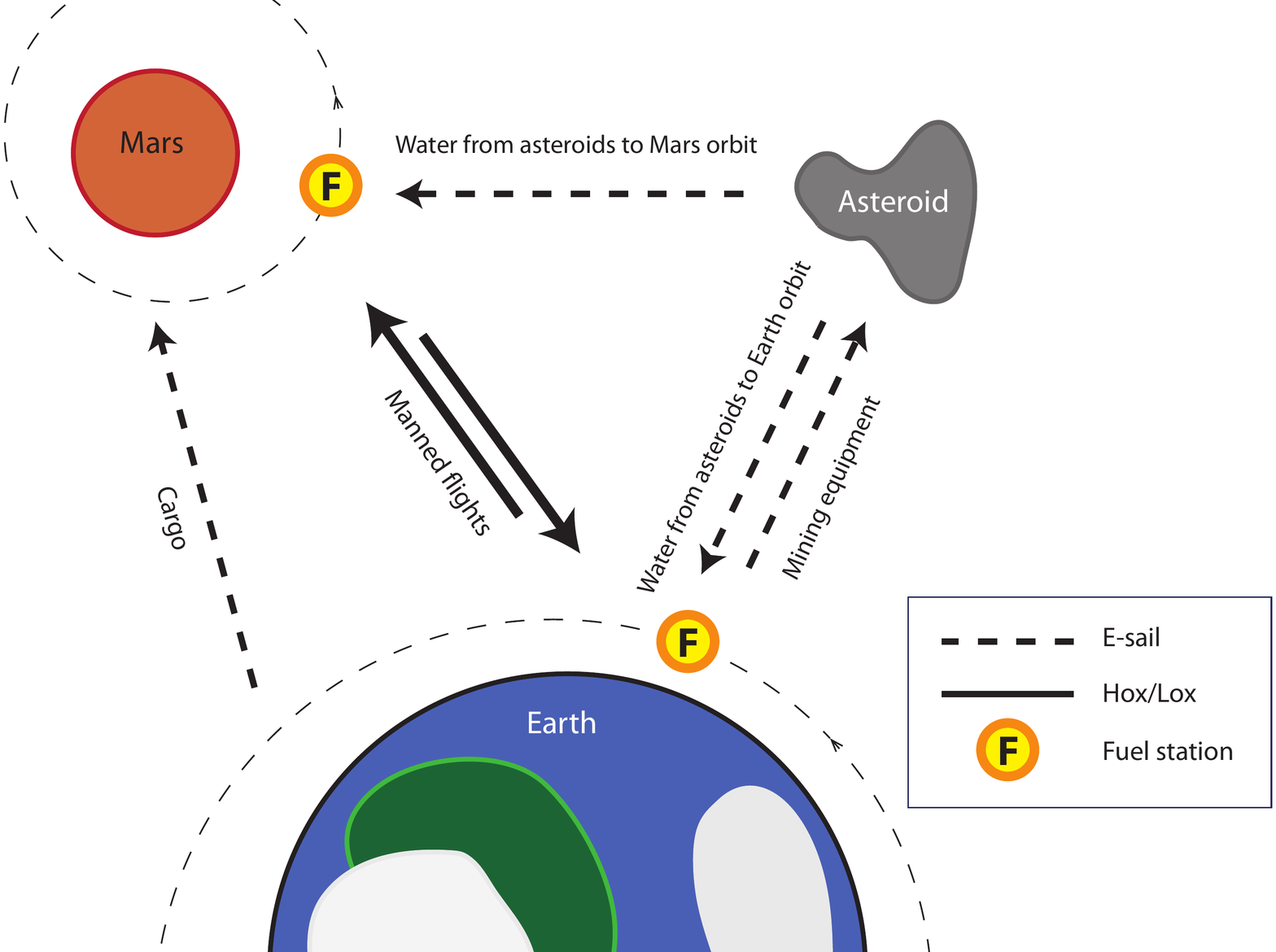}
 \caption{Outline of EMMI, showing the locations of the orbital fuel stations where the water is split and condensed into LOX/LH2 fuel. On these fuel stations manned vehicles traveling between Earth and Mars can be fueled, dramatically reducing the overall mission fuel ratio at launch. Cargo transport and mining activities take place propellantlessly using E-sails.}
 \label{fig-emmikuva}
 \end{figure}

The alternative technological path that we propose in this paper could provide means for affordable and continuous trafficking of cargo and passengers between the Earth and Mars and thus continuous presence of human beings on the surface of Mars. Because of the exponential nature of the rocket equation, being able to refill the fuel tanks on the way one could decrease the transportation cost tremendously. This can be achieved by hauling water from the asteroids and converting it to LOX/LH2 fuel in orbital fueling stations.  An outline of this scheme is shown in Fig.~\ref{fig-emmikuva}. Other assets required by our proposed E-sail facilitated manned Mars presence include the possibility to use the water as a radiation shield, potable water and a source of breathable oxygen. We call the scheme presented on this paper the E-sail facilitated Manned Mars Initiative (EMMI).

The purpose of this paper is to analyze EMMI schemes at high level of abstraction by attempting to identify the leading terms (drivers) of their mass and risk budget. In particular cases, we shall also dwell on some details, but full engineering design of the needed space assets is outside the scope of the paper.

\section{Electric solar wind sail}

The electric solar wind sail (E-sail) provides thrust in the solar wind without consuming fuel~\cite{janhunen2004, janhunen2010big}. The E-sail uses charged tethers to extract momentum from the solar wind ions (mainly protons) by electrostatic Coulomb interaction with the plasma flow to produce thrust. According to current estimates, the E-sail is 2-3 orders of magnitude more efficient than traditional propulsion methods (chemical rockets and ion engines) in terms of produced lifetime-integrated impulse per propulsion system mass, if the mission duration is 10 years and the mission does not proceed too far into the outer solar system \cite{janhunen2010big}. This is based on numerical simulations \cite{janh2009}. When these simulations are run for plasma representing LEO conditions \cite{JanhunenKoln2014}, their predicted electron sheath width (which is a proxy for E-sail thrust per length) is in good agreement with laboratory measurements of the sheath width in LEO-like conditions \cite{SiguierEtAl2013}. The thrust of an E-sail is inversely proportional to the distance from the Sun as $F \sim 1/r$ \cite{janh2009}. In comparison, the thrust produced by a photonic solar sail decays faster as $F \sim 1/r^2$. Thus the E-sail has particular potential for long-lasting solar system missions.  The E-sail requires no fuel and the charging of the wires can be accomplished by an electron gun powered by modest-sized solar panels. The electric power consumption per produced thrust is 1 kW/N at 1 au from the Sun \cite{janhunen2013}. For fixed tether voltage, the power consumption is proportional to the electron current gathered by the tethers, which is in turn proportional to solar wind density, behaving as $\sim 1/r^2$. Because this is the same scaling as the illumination of the solar panels, it implies that if the solar panels are dimensioned to power the E-sail at a specific distance such as 1 au, they also suffice to power the sail at other distances as well if one assumes constant panel efficiency.

To enable maneuvering and trajectory control, the E-sail thrust can be vectored in a cone of $\sim 30^\circ$ around the solar wind velocity vector~\cite{spin2013}. It is possible to adjust the magnitude of the thrust between 0-100\,\% by modifying the current and voltage of the electron gun. Turning E-sail propulsion off is possible at any time by turning off the electron gun. A strategy of maximizing available thrust by matching the electron gun current with the current gathered by the tethers leads by certain natural negative feedback mechanisms to a situation where the thrust varies much less than the solar wind dynamic pressure \cite{toivanen_janhunen2009}. Even with simple trajectory control law, the maneuverability of the E-sail is sufficient to allow navigation to, for example, Mars \cite{toivanen_janhunen2009}.

Applications of the E-sail have been extensively researched~\cite{janhunen2014}, outlining for example outer planet exploration~\cite{janhunen2013uranus, quarta2010, mengali2008}, asteroid flyby~\cite{mengali2014}, asteroid rendezvous~\cite{quarta2010b}, NEO sample return~\cite{quarta2014}, and hazardous asteroid deflection~\cite{merikallio2010}.



\section{Asteroid mining}
At the core of the E-sail facilitated Manned Mars Initiative (EMMI) is the utilization of the propellantless thrust provided by the E-sail to haul water from asteroids into orbits of Earth and Mars. A large E-sail, that can provide 1 N of thrust at 1 au from Sun, can travel from the Earth to the asteroid belt in a year, and bring back three tonnes of water in three years~\cite{quarta2010b, quarta2014}. 
One E-sail based spacecraft is capable of repeating the journey multiple times within its estimated lifetime of at least ten years. 

\subsection{Extracting water from asteroids}
We outline here a baseline scheme for extracting water from the asteroids. An alternative method will be discussed in the following section.
In this section we introduce a twin container, of which the other part is heated with electric power generated by solar panels. The evaporating water vapor is condensed in the other container.

\begin{figure}
 \noindent\includegraphics[width=15cm]{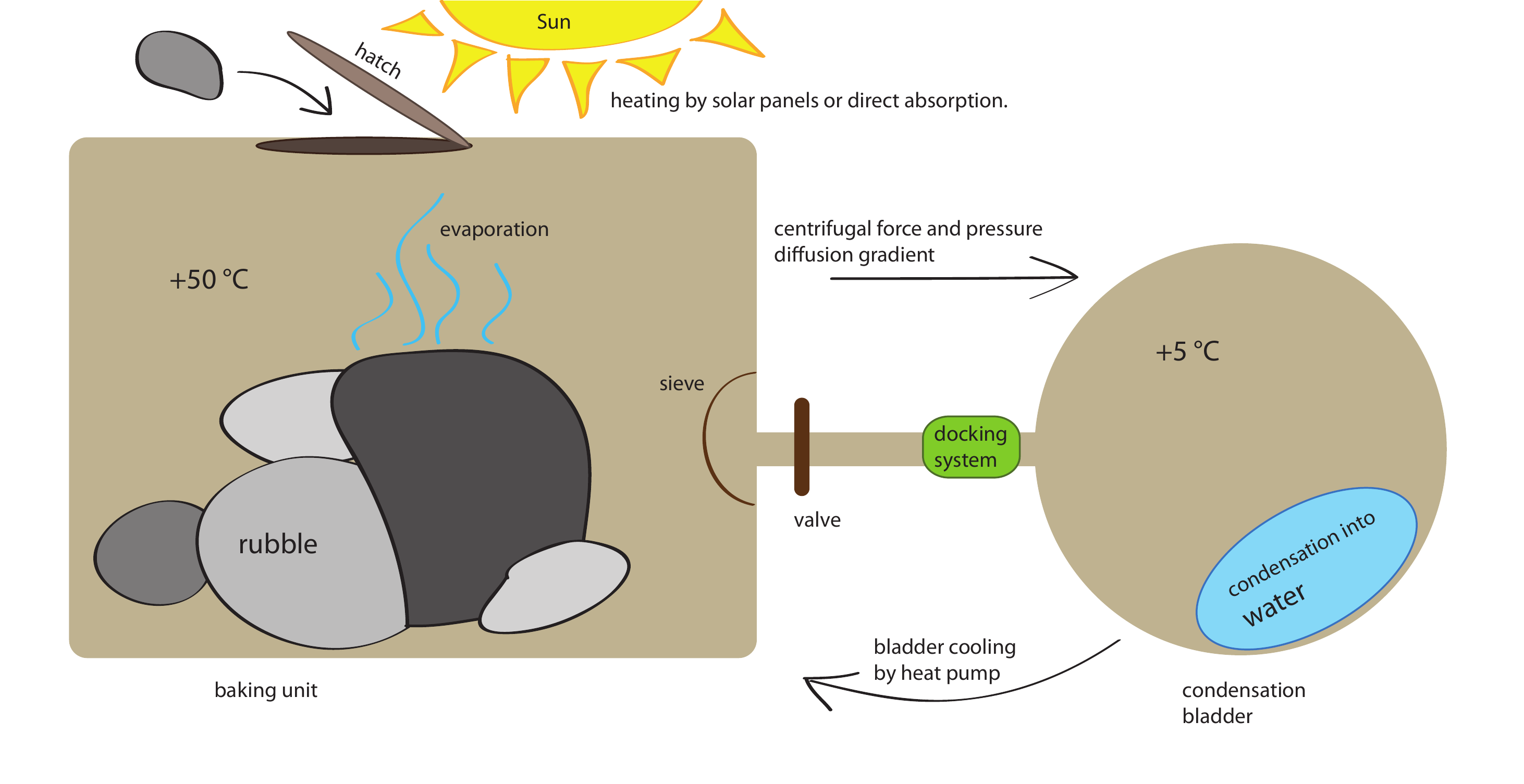}
 \caption{A schematic presentation of the water extractor unit.}
 \label{fig-extractor}
 \end{figure}

The extractor apparatus consists of two container parts connected with a pipe (Fig.~\ref{fig-extractor}). The asteroid material is gathered into the other container, which we call an oven from here on. The oven is closed up and tightly sealed, after which it is heated up to about +50$^o$C for the water contained in the asteroid material to evaporate 
(at +50$^o$C, water boils at 0.13 bar). The water vapor thus fills the oven and starts flowing through the connecting pipe to the other container (the tank). Temperature of the tank is held slightly above freezing, say at +5 $^o$C, which causes the water vapor to condense on the walls of the tank, but keeps it from freezing and blocking the connecting pipe. After the flow of the vapor recedes, the valve in the connecting pipe is closed and the asteroid material of the oven can be exchanged for another batch of asteroid material. Once the tank is filled up to the desired level, it can be disconnected from the connecting pipe and transported to its final destination. A new, empty tank can be connected to the oven and the process repeated for as long as there are empty tank modules available. These can be transported en masse from Earth to the asteroids with an E-sail.

A semipermeable membrane that only lets the water pass in vapor form could be installed on the connecting pipe between the oven and the tank to ensure that the water stays in the tank. This membrane could be heated to keep any larger water mass from forming on top of it (and thus blocking the vapor transfer). Alternatively, the gravity of the asteroid might be enough to help guiding the condensing water towards the end of the tank. The gravity could also be of an artificial (rotational) origin, which could be arranged if the whole extractor apparatus is lifted from the surface of the asteroid and made to spin in space, as is discussed in the section~\ref{sec-alternativewater}.

To get a handle on the mass and cost of asteroid mining, let us estimate its energy consumption. The ice-fraction of the asteroid requires 2900 kJ/kg to melt and evaporate into +50$^o$C steam, if an initial temperature of -50$^o$C for the asteroid ice is assumed. Using a heat pump [assuming coefficient of power (COP) of 3] for transferring the condensation heat from the tank in to the oven, we could reduce the energy demand down to 1300 kJ/kg.
However, asteroids mainly consist of rocky material, which has to be heated in the process as well, here from the initial -50$^o$C up to the +50$^o$C. 
Assuming the target asteroid to have water content of 10\%, rest 90\% being basalt (c$_p$ = 0.84 kg kJ$^{-1}$K$^{-1}$), the total process would require 2000 kJ/kg. In the case of a dry asteroid with the water content of barely 2\%, the energy required to extract a kilogram of water would raise up to 5400 kJ/kg. Moreover, on the surface of Mars the combination of only 2\% of soil water content and of that water being released only when the soil is heated up to around 400$^o$C increases the energy requirement up to 20 MJ/kg~\cite{curiosity2013}.

We assume that we can find and choose an asteroid with the desired high (10\%) water content, which leads into baseline electric energy requirement for water extraction of 2 MJ/kg.

\subsection{Alternative approach to water extraction}
\label{sec-alternativewater}
Another, perhaps more elegant way to extract water from the asteroid material would be to adjust the container temperature by alternating the containers surface albedo and thus by using only direct solar energy. Here the oven would need to be coated by a material whose optical absorptivity is much higher than the infrared emissivity, such as gold or copper \cite{gilmore2002}, and the tank cooled down by a cold coating, e.g.~white paint. The oven would then be filled with asteroid material, or a small asteroid could even be enveloped in its entirety (in the way of recently proposed NASA Asteroid Initiative~\cite{keck2012}), lifted from the surface and set into a rotating motion. Rotation would ensure that the tank does not shadow the oven for prolonged periods, and more importantly, it creates an artificial gravity keeping the condensed water at the outer wall of the tank.
Once the extraction is complete, the oven and the tank would be separated, the oven discarded and the now ice-filled tank left to wait for its carrier E-sail, constantly beaconing to ensure it will not get lost.


\subsection{Transportation of water by E-sails}

\begin{figure}
 \noindent\includegraphics[width=10cm]{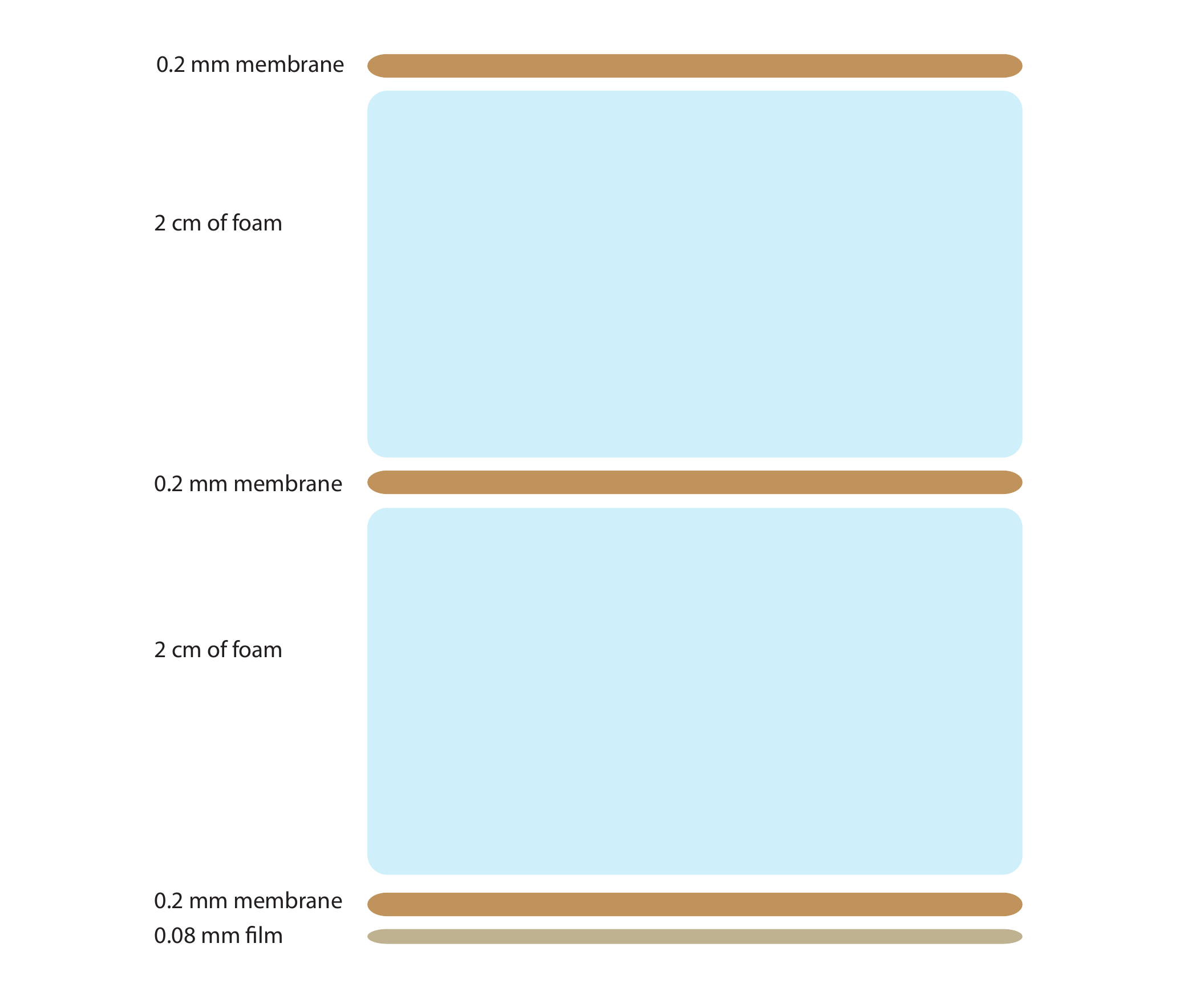}
 \caption{A schematic drawing showing the layers of the tank wall.	Multiple layers and a space filler in-between them are necessary preparations for micrometeorite impacts.}
\label{fig-tank}
 \end{figure}

Transporting the liquefied water from the asteroids first to the heliocentric orbits of the Earth and Mars and then to bound orbits could be done in lightweight, thin walled container tanks, each weighing a few tonnes when filled. 
The container 
has to be specially designed to endure the space environment. The main requirements are containment ability, low mass, resistance to micrometeoroids and tolerance of possible freezing/thawing cycles of the cargo. For these purposes, we propose a layered membrane structure portrayed in Figure~\ref{fig-tank}. 

The cargo could be brought from a high orbit down to LEO without fuel consumption by using gradual aerobraking. The thin walled container tank conducts heat to the water mass it is carrying, thus making the tank heat resistant. During aerobraking, the tank could be maneuvered to be in the front so that all the control electronics could be protected behind it.

\subsection{Fuel from water} 
\label{sec-fuel}
The water, once transported to orbits of Mars and Earth, is split into its constituent hydrogen and oxygen by electrolysis or by photocatalytic means~\cite{nature2013}. After liquefaction, one obtains cryogenic rocket fuel. EMMI is based on manufacturing this LOX/LH2 fuel from the water transported from the asteroids. With 20 kW of electric power one can produce 30 000 kg of LOX/LH2 from water in a year.

However, if liquefying H$_2$ in temperatures prevalent close to asteroids proves difficult, producing an methane-oxygen fuel could be an alternative possibility. This type of fuel is more complex to produce from asteroid material. Water would also now be split into oxygen and hydrogen, but the pure oxygen would be used to burn carbon containing asteroid material into $CO_2$. The carbon dioxide would then be combined with hydrogen in a Sabatier reaction to produce methane and water ($CO_2 + 4 H_2 \rightarrow CH_4 + 2 H_2O$).

Water to fuel conversion plants would reside in high orbits, i.e. barely captured in the gravity well of Earth or Mars. Storages of fuel would be accumulated on these strategic places well in time before they are needed. The redundancy can be built into the system, having a number of smaller tanking stations with combined storage capacity exceeding the required mission fuel consumption. Thus malfunctioning of any one fuel production station would not be mission critical. Fuel would also be produced so that it would be ready and waiting before the first humans are sent to space.



\section{Operations on Mars}
\label{sec-operationsMars}
It has been estimated from the Curiosity Mars rovers measurements that around $1.5\% - 3\%$ of the mass of Martian surface soil is water that can be released by heating the material up to $200^\circ$C $- 400^\circ $C~\cite{curiosity2013}. This water, once extracted, could be split into same kind of fuel to that produced from the water mined from the asteroids, see sec.~\ref{sec-fuel}, thus allowing to use the same rocket engines (and thus the same vehicles) for lift-off from Mars as for other parts of the mission. This could dramatically further reduce the costs of the mission, albeit also posing additional design restrictions for the spacecraft.

Landing humans on Mars could be circumvented altogether by using remote controlled robotics as proposed in HERRO concept (Human Exploration using Real-time Robotic Operations, \cite{herro2012}): instead of landing, humans would orbit around Mars and teleoperate robots on its surface. This approach would significantly reduce the costs and risks and alleviate the issue of forward contamination of Mars by human-carried microbes.

\section{Mission outline}

EMMI 
proposes means for continuous habitation on the Martian surface. Recurrent trips with LOX/LH2 powered transport vehicle between Earth and Mars would transport the exchange crews and their food in regular intervals. Additional support equipment, food and accessories could be taxied from the Earth propellantlessly with E-sails. 
The crew transport vehicles would be fueled up on the fuel stations situated in high planetary orbits and/or planet-Sun Lagrange points. The $\Delta$V required for LEO -- Lagrange point transfer and for the Lagrange point -- Mars transfer is in the scale of 2.5-3.3 km/s, meaning the fuel requirement is about 50$\%$ - 60$\%$ of the wet mass of the spacecraft if LOX/LH2 is used, moderate Isp of 420 s is assumed and 10\% fuel margin is used.

The mission could thus be done on separate stages, filling up the fuel tank in-between each leg. First, a launcher is used to lift the passengers/payload to LEO where the first tanking occurs. Alternatively, the lift is continued with electric propulsion engines through the magnetosphere. The craft is lifted to L1, L2 or a high orbit, where again fuelling from the awaiting fuel reservoirs occurs. The last leg from Earth orbit to Mars and capture to Mars orbit would again consume the fuel.

The target asteroids would be sought from the vicinity of the Martian orbit, at around 1.5 au, as this is where majority of the mining products are headed.
Yearly water extraction pace of roughly 50 000 kg would suffice for a manned bidirectional trip between Earth and Mars taking place every other year.
To achieve this, 3.2 kW of electric power on the asteroid is  required. At the distance of 1.5 au from the Sun, this translates into 230 kg of solar panels assuming a horizontal panel on the equatorial surface of a rotating asteroid and a characteristic mass for the solar panels of 100 W/kg at 1 au (Joel Ponzy, private communication). As power is also needed for other purposes, such as moving, communications and countering power system aging, we will assume a 50\% higher power consumption, thus arriving at the whole extractor power system weighing 340 kg. We estimate that the whole extraction vessel would in this case weigh around 2000 kg, which is much lower than the expected mass of the orbital fuel factories. This and other key mass figures of EMMI are listed in Table~\ref{table-1}.

For the spacecraft carrying the astronauts there is a need for radiation shielding, which increases the mass of the manned vehicle considerably. The water stored on fuel stations can be used as a radiation shield, thus removing the need to launch a heavy shielded manned module all the way from the Earth's surface. Potable water for the needs of the crew and oxygen for breathing could also be manufactured from this water.


\begin{table}
\caption{Some key mass figures of EMMI.}
\centering
\label{table-1}
\begin{tabular}{l r l}
\noalign{\smallskip}\hline
Production and transportation of water per year  & 50 000 & kg\\
extractor vehicle weight & 2000 &  kg\\
Transporter E-sail & 500 & kg\\
Manned vehicle & 50 000 & kg \\
Payload transferred from Earth to Mars & 10 000 & kg yearly \\
\end{tabular}
\end{table}

\subsection{Manned transfer vehicle}
The vehicle transporting the crew between Earth and Mars orbits would have a mass of 50 000 kg, including radiation shielding but not including fuel.
As sturdy radiation protection is mandatory on manned Mars flights~\cite{zeitlin2013, hellweg2007, cucinota2012}, 40\% of the mass (20 000 kg) would be radiation shielding water, gotten on-board from Earth's orbital tanking station. Only the crew and their food is needed to be launched from Earth by traditional means; the transfer vehicle would be waiting on a high orbit around Earth with a full tank of LOX/LH2.

The manned transfer vehicle would shuttle in-between orbits of Earth and Mars, getting fueled from the LOX/LH2 produced from the asteroid water at the tanking stations residing on these locations. 
\subsection{Drop to Mars - alternative landing method}



The normal way of landing on Mars requires the use of a heat shield to combat the excessive temperatures generated during a high-speed entry into the atmosphere. Given low cost fuel in Mars orbit, however, one could also consider an all-propulsive landing reducing the speed of the spacecraft before it enters the atmosphere. The craft would fill its propellant tanks once more in orbit around Mars, make a large burn with its engine to almost cancel its orbital speed and then drop into the atmosphere with a speed much lower than that experienced when following traditional entry trajectories. This together with a propulsive insertion into orbit around Mars would diminish the mass fraction of the vehicle's thermal protection system close to zero. 

An all propulsive descent would also make it easier to land precisely as it allows greater control over the trajectory. Only a short trajectory through the atmosphere is necessary which reduces the possibility of navigation errors propagating.
In a partly similar concept, an idea of reducing heat shield demands by using propulsive landing to Mars was introduced by \cite{Marsh2011}.

Control over the landing point is an essential aspect once a colony has been built, as landing too near or too far from it could be hazardous. A possibility exists to design the crew transport vehicle, which transports humans between Earth and Mars, to also function on transfers in-between Martian surface and orbit. This would, however, increase the design requirements of the spacecraft.

\subsection{EMMI timeline}
EMMI would first start with launching of asteroid mapping E-sails to scout for suitable water and/or carbon containing asteroids. After the mapping phase (duration approximately four years), the mining spacecraft would be sent to selected target asteroid and the mining would start (1 year to transport, 1.5 years to set up operations and to produce the first 50 000 kg of water). A third of the mined water would be transferred to Earth's and two thirds to Mars orbit. After 8 years from the first launch, one would have 30 000 kg of fuel on the orbit of Earth. 

At the first stage of operations, humans would stay on the orbit of the Mars, letting HERRO~\cite{herro2012} take care of the surface operations. This would greatly reduce the costs at this stage as the astronaut habitat, life support and return launch from the surface of Mars would not need to be considered. However, HERRO systems would facilitate the mapping of the surface and robotically prepare the base for the future crews that could then land on the surface more prepared.

As EMMI continues, mining operations are expanded to additional asteroids to provide a more continuous and reliable fuel supply. Redundancy can be built up by additional fuel stations on strategic locations. Continuous human presence on the surface and/or on orbit of the Mars would be possible with affordable costs and symmetrically bidirectional traffic.

Once the asteroid mining and fuel transportation system is built, its maintenance would be relatively low cost as all of the craft included are reusable and all the advantages from serial production (lower production and designing costs per unit) could be taken advantage of. Thus only crew, spare parts, and life support material has to be put into the system and launched into orbit analogously to the maintenance of the International Space Station (ISS), currently orbiting the Earth with continuous manned presence \cite{nasa2014}. This makes the maintenance effort and thus also the costs between the EMMI and ISS roughly comparable with each other for each astronaut-year.

\section{Development risks}

EMMI is based on providing water in Earth and Mars orbits, which in turn is based on E-sail propulsion and on asteroid water mining. These technologies have not yet been demonstrated at high TRL and therefore the EMMI scheme as a whole must be considered at least somewhat risky at the present time. However, demonstration of asteroid water mining and its transportation by E-sail propulsion to planetary orbit is an easily scalable process. This means that it can be first demonstrated in a small scale at low cost (many orders of magnitude lower than full EMMI) and then scaled up. Manufacturing cryogenic fuel in orbit is also something that has not yet been done, but we think that few would doubt its feasibility. Thus, demonstration of the E-sail and small-scale asteroid water mining would strongly reduce uncertainty about EMMI's feasibility, and the investment needed would be vanishingly small compared to the cost of manned space activity.

\section{Summary and conclusions}
We analyzed how E-sails could act as critical enabling technology for setting up continuous manned bidirectional traffic to Mars, using asteroid water mining and orbital fuel manufacturing. The most massive and therefore the most costly part of the fuel supply chain are the orbital fuel factories, because splitting water into hydrogen and oxygen takes much more energy than liberating it from asteroid soil. The E-sails are a relatively minor part of the total mass budget. The efficiency of the overall concept stems from the intermediate tankings that, due to the exponential nature of the rocket equation, reduce the fuel consumption considerably compared to the traditional approach without refueling opportunities. 

It is also a significant cost benefit that the orbital water can be used as crew radiation shield. We think that with EMMI continuous manned presence on Mars (with bidirectional traffic) would be possible at a cost level which is comparable to that of maintaining the International Space Station, ISS.


\section{Acknowledgements}
The work was partly funded by Academy of Finland, grant number 250591.

\section{Bibliography}








\end{document}